\documentclass[preprintnumbers,twocolumn,prl,aps,groupedaddress,footinbib,amsfonts,amsmath,amssymb,showpacs]{revtex4-1}

\usepackage[T1]{fontenc}
\usepackage{blindtext}

\usepackage{graphics,graphicx}
\usepackage{dcolumn}
\usepackage{bm}
\usepackage{epsfig}
\usepackage[usenames]{color}
\usepackage[hidelinks]{hyperref} 
\usepackage{mathbbol}
\usepackage{epstopdf}
\usepackage{simplewick}
\usepackage[utf8]{inputenc} 
\usepackage{slashed}
\usepackage{soul}
\usepackage[dvipsnames]{xcolor}
\usepackage[export]{adjustbox}
\usepackage{caption}
\usepackage{multirow}

\usepackage{lineno} 

\usepackage{hyperref} 
\hypersetup{
    colorlinks,
    citecolor=black,
    filecolor=black,
    linkcolor=black,
    urlcolor=black
}

\begin{document} 


\title{Transformation of transverse momentum distributions from Parton Branching to Collins-Soper-Sterman framework}



\author{Armando~Bermudez~Martinez}\email{armando.bermudez.martinez@cern.ch}

\affiliation{CERN - 1211 Geneva 23 - Switzerland}

\begin{abstract}
\noindent
Two main frameworks for defining transverse momentum dependent (TMD) parton densities are the Collins-Soper-Sterman (CSS) formalism, and the Parton Branching (PB) approach. While PB-TMDs have an explicit dependence on a single scale which is used to evolve PB-TMDs in momentum space, TMDs defined in CSS formalism present a double-scale evolution in renormalization and rapidity scales, via a pair of coupled evolution equations. In this letter I leverage the Collins-Soper kernel determined from simulated Drell Yan transverse momentum spectra using PB-TMDs, and provide, for the first time, the transformation of TMD parton distributions from the PB framework to the CSS formalism. The evolved PB-TMDs in $b$-space are compared to the recently released, unpolarized TMD distribution ART23.
\end{abstract}

\maketitle

{\bf Introduction.} The production of colorless final states at a reference scale $\mu$ in high-energy hadron collisions is described by the factorization~\cite{Collins:1989gx} of perturbative short distance scattering cross-sections and non-perturbative long-distance parton distribution functions (PDFs), given $\mu \gg \Lambda_\text{QCD}$. When describing the transverse momentum spectra of a colorless final state, additional non-perturbative contributions need to be considered besides the PDFs. These contributions are the result of the intrinsic transverse momentum motion of the colliding partons, and also of non-perturbative components of Sudakov form factors that resum soft radiation. Two of the main frameworks which account for these effects are the Collins-Soper-Sterman (CSS) formalism based on transverse momentum dependent (TMD) factorization~\cite{Collins:2011zzd, Echevarria:2011epo, Becher:2010tm}, and the Parton Branching (PB) approach~\cite{Hautmann:2017fcj,Hautmann:2017xtx}. In this work I examine the connection between these two seemingly unrelated formalisms and provide for the first time the transformation of TMDs from the PB to the CSS framwork. In order to evolve the PB-TMDs with the CSS evolution equations two main ingredients are needed, the starting distribution and the non-perturbative Sudakov factor defined by the rapidity anomalous dimension also called Collins-Soper (CS) kernel. Since the CS kernel is not an explicitely defined in the PB method I determine it using the method proposed in~\cite{BermudezMartinez:2022ctj} from Drell-Yan (DY) transverse momenta spectra. 

\textbf{The PB approach} provides an evolution equation for transverse momentum dependent parton densities ${\cal A}_a$, which has the integral form:

\begin{align}
    & {\cal A}_a (x, k_{\bot}^2; \mu^2) = {\cal A}_a (x, k_{\bot}^2)\Delta_a(\mu^2, \mu_0^2) \nonumber \\ 
    {}& + \int\frac{d^2\mu_\bot^{\prime}}{\pi\mu_\bot^{\prime 2}} \Delta_a(\mu^2, \mu_\bot^{\prime 2})\Theta(\mu^2-\mu_\bot^{\prime 2})\Theta(\mu_\bot^{\prime 2}-\mu_0^2)\nonumber \nonumber \\ 
    {}& \times  \sum_b \int_x^{z_M} dz P_{ab}^R(z; \alpha_s){\cal A}_b\left( \frac{x}{z}, (k_{\bot}+(1-z)\mu_\bot^{\prime})^2; \mu_\bot^{\prime 2}\right)\;, \label{eq1}
\end{align}
where $x$ is the longitudinal momentum fraction of parton, $k_{\bot}$ its transverse momentum, and $z$ the transfer of the longitudinal momentum from parton of flavor $b$ to parton of flavor $a$. In addition $P_{ab}^R$ are the real emission part of the DGLAP splitting functions~\cite{Gribov:1972ri, Lipatov:1974qm, Altarelli:1977zs, Dokshitzer:1977sg}, and $\alpha_s$ is the strong coupling. The no-emission probability $\Delta_a(\mu_2^2, \mu_1^2)$, also called Sudakov form factor, between two scales $\mu_1$ and $\mu_2$ is defined as:
\begin{align}
& \Delta_a(\mu_2^2, \mu_1^2) = \nonumber \\
& \exp\left( -\sum_b\int_{\mu_1^2}^{\mu_2^2}\frac{\textrm{d}\mu^{\prime 2}}{\mu^{\prime 2}}  \int_0^{z_M} \textrm{d}z\; zP_{ba}^R(z;\alpha_s) \right)\;,
\label{eq:RealSud}
\end{align}
where $z_M$ is the soft-gluon resolution scale which separates resolvable from non-resolvable emissions. 

The starting distribution ${\cal A}_a (x, k_{\bot}^2)$ in eq.~\ref{eq1} parametrizes the parton distribution at the starting scale $\mu_0$ and it is factorized in a collinear part, and in a non-perturbative (NP) transverse momentum dependent function $f_{NP}$ as: 
\begin{equation}
    \mathcal{A}_a(x, k_{\bot}^2)= f_a(x,\mu_0)\cdot g_{NP}(k_{\bot}^2),
    \label{eq3}
\end{equation}
where $f_a(x,\mu_0)$ is the integrated TMD, while the function $g_{NP}$ is the intrinsic transverse momentum distribution usually modelled as a Gaussian function with a parametrized width. The PB method provides a good description of the DY transverse momentum spectrum in a very wide range of DY masses and center-of-mass energies as shown in~\cite{BermudezMartinez:2019anj, BermudezMartinez:2020tys, Martinez:2021mzy, CMS:2019raw}. It has also been shown to support TMD factorization at low transverse momentum~\cite{BermudezMartinez:2022ctj}, and in addition it can provide a good description at large transverse momentum and high jet multiplicity via multi-jet merging using the TMD merging algorithm~\cite{Martinez:2021chk,BermudezMartinez:2022bpj}.

In this work the PBset2 PB-TMD~\cite{BermudezMartinez:2018fsv} is employed. It is worth noting that in the PB approach the collinear, integrated TMD density is fitted to collider data, which for the case of PBset2 corresponds to HERA I+II inclusive DIS cross section measurements~\cite{H1:2015ubc}.

\textbf{The CSS factorization} provides the evolution of TMD distributions via a pair of equations:
\begin{eqnarray}
    \mu^2 \frac{d}{d\mu^2} F(x, b;\mu, \zeta) &=& \frac{\gamma_F(\mu, \zeta)}{2} F(x, b;\mu, \zeta), \label{eq4}\\
    \zeta \frac{d}{d\zeta} F(x, b;\mu, \zeta) &=& - {\cal D}(\mu, b) F(x, b; \mu, \zeta), \label{eq5}
\end{eqnarray}
where $F$ is the TMD parton distribution which depends on the parton longitudinal momentum fraction $x$ and transverse distance $b$. The evolution variables $\mu$ and $\zeta$ arise from the renormalization of the ultraviolet divergences and from the factorization of rapidity divergences respectively. The function $\gamma_F$ is the TMD anomalous dimension and the function ${\cal D}(\mu, b)$ is the rapidity anomalous dimension, also called CS kernel. The solution of eqs.~\ref{eq4} and~\ref{eq5} for a given flavor and starting distribution $F(x, b)$ can be expressed as~\cite{Scimemi:2018xaf}.
\begin{equation}
    F(x, b; \mu, \zeta) = R [b; (\mu, \zeta) \rightarrow (\mu_0, \zeta_0)] F(x, b),
    \label{eq6}
\end{equation} 
where $R$ is the evolution factor along a path in the $(\mu, \zeta)$ plane. Here I use the $\zeta$-prescription~\cite{Scimemi:2019cmh, Scimemi:2018xaf}, which decorrelates the TMD distribution and the CS kernel. Within this framework, unpolarized TMD parton distributions have been determined from global fit analyses of vector boson production and Semi-Inclusive Deep-Inelastic scattering data~\cite{Scimemi:2017etj,Bacchetta:2019sam, Bertone:2019nxa, Scimemi:2019cmh, Bacchetta:2022awv, Moos:2023yfa}. The parameters determined in these fits correspond to the starting TMD distribution and the CS kernel.  

In contrast to the PB method, collinear non-perturbative effects encoded in the PDFs are not fitted in this framework. Instead, available PDF global fits are used like the ones provided by the HERA~\cite{H1:2015ubc}, NNPDF~\cite{NNPDF:2017mvq}, CTEQ~\cite{Hou:2019efy}, and MSHT~\cite{Bailey:2020ooq} collaborations.
In~\cite{Bury:2022czx} a systematic investigation of the role of PDF bias in TMD determinations was performed.

For the purpose of comparison in this work employ the recently released ART23 TMD distribution~\cite{Moos:2023yfa}, which includes the Z- and W-boson production data, and uses the MSHT20 PDF~\cite{Bailey:2020ooq} as base collinear distribution. 

{\bf Evolution of PB-TMDs using the CS kernel.} The main difference between the evolution defined in eq.~\ref{eq1} compared to eqs.~\ref{eq4},~\ref{eq5} lies on the rapidity scale evolution in eq.~\ref{eq5}. The CS kernel governs the evolution in rapidity scale, and contains information on long-range forces acting on quarks~\cite{Vladimirov:2020umg}. Even if not explicitely defined, the CS kernel underlying the PB approach can be determined from cross-sections, without any reference to the underlying TMD distributions, as shown in~\cite{BermudezMartinez:2022ctj}. The main result of this letter is the use of the determined CS kernel to evolve the PB-TMD starting distribution. In this manner, the PB-TMDs are expressed in the CSS formalism for the first time, a long standing problem in the TMD community. It is worth pointing out that this correspondance is possible in the $\zeta$-prescription because in this prescription the notion of modeling of the TMD distribution, and the influence of the TMD evolution are disentangled~\cite{Scimemi:2018xaf}.

In the asymptotic $b \rightarrow 0$ limit, the operator product expansion (OPE) of the TMD distribution allows to construct a phenomenological anzats~\cite{Scimemi:2019cmh} connecting the collinear part of the TMD distribution with a non-perturbative function dependent on the transverse distance. The starting distribution of a parton of flavor $a$ can then be written as:
\begin{align}
    & F_a(x, b) = \int_x^1 \frac{dy}{y} \sum_{b}C_{a\leftarrow b}(\frac{x}{y}, \textbf{L}_{\mu_\text{OPE}}, \alpha_s)    \nonumber \\
    & \times f_{b} (y, \mu_\text{OPE}) f_{NP}(x, b),
    \label{eq7}
\end{align}
where $C$ are the matching Wilson coefficients, $\alpha_s$ is the strong coupling evaluated at the OPE scale $\mu_\text{OPE}$, and $\textbf{L}_{\mu} = \ln(b^2\mu^2/(4\exp(-2\gamma_E)))$, with $\gamma_E$ being the the Euler constant. The scale $\mu_\text{OPE}$ is chosen such that it minimizes the logarithmic contribution at $b\rightarrow 0$, and does not reach the Landau pole. Similar to~\cite{Scimemi:2019cmh} I use the relation:
\begin{equation}
    \mu_\text{OPE} = \frac{2e^{\gamma_E}}{b} + \mu_0,
    \label{eq8}
\end{equation}
where $\mu_0 = \sqrt{1.4~\text{GeV}^2}$ corresponds to the reference scale of the integrated TMD for the case of PBset2~\cite{BermudezMartinez:2018fsv}. This value of $\mu_0$ provided the best fit result for PBset2~\cite{BermudezMartinez:2018fsv}. The coefficient functions $C$ are known up to next-to-next-to-leading-order~\cite{Gehrmann:2014yya, Echevarria:2015usa, Echevarria:2016scs, Luo:2019hmp}. For simplicity I use the leading-order (LO) expression~\cite{Echevarria:2016scs, Collins:2014jpa} $C^{[0]}_{a\leftarrow b}(x) = \delta_{ab} \delta(1-x)$ in eq.~\ref{eq7}, which for a given flavor results in:
\begin{equation}
    F_a(x, b) = f_a (x, \mu_\text{OPE}) f_{NP}(x, b) ,
    \label{eq9}
\end{equation}
It is worth noting that when the coefficient functions are considered at LO, the functional structure for the starting TMD distribution is equivalent to that of the PB approach given in eq.~\ref{eq3}, where $g_{NP}(k_{\bot}^2)$ corresponds to the Hankel transformation of $f_{NP}(b)$.

\begin{figure}[h]
\includegraphics[width=0.495\textwidth]{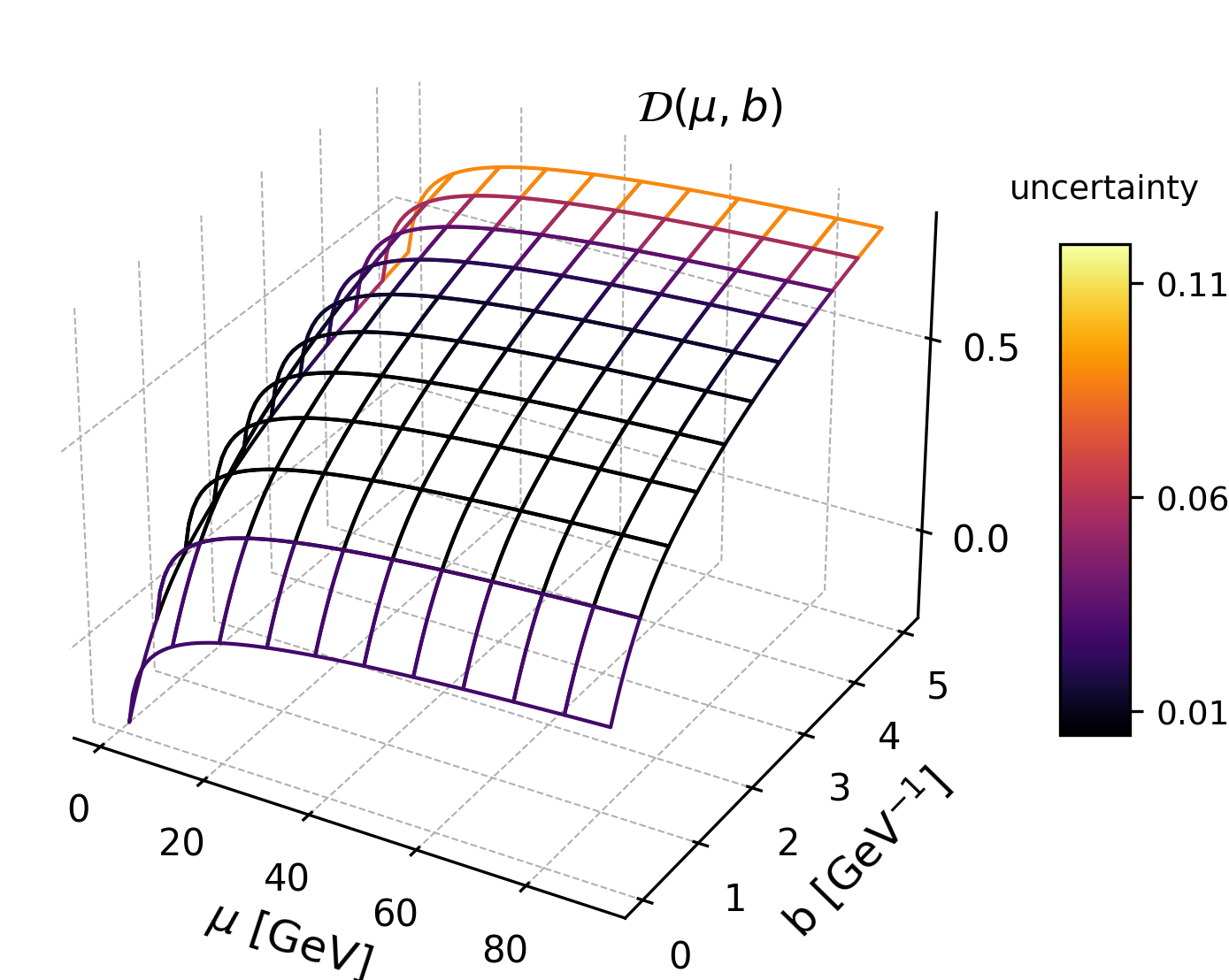}
\includegraphics[width=0.495\textwidth]{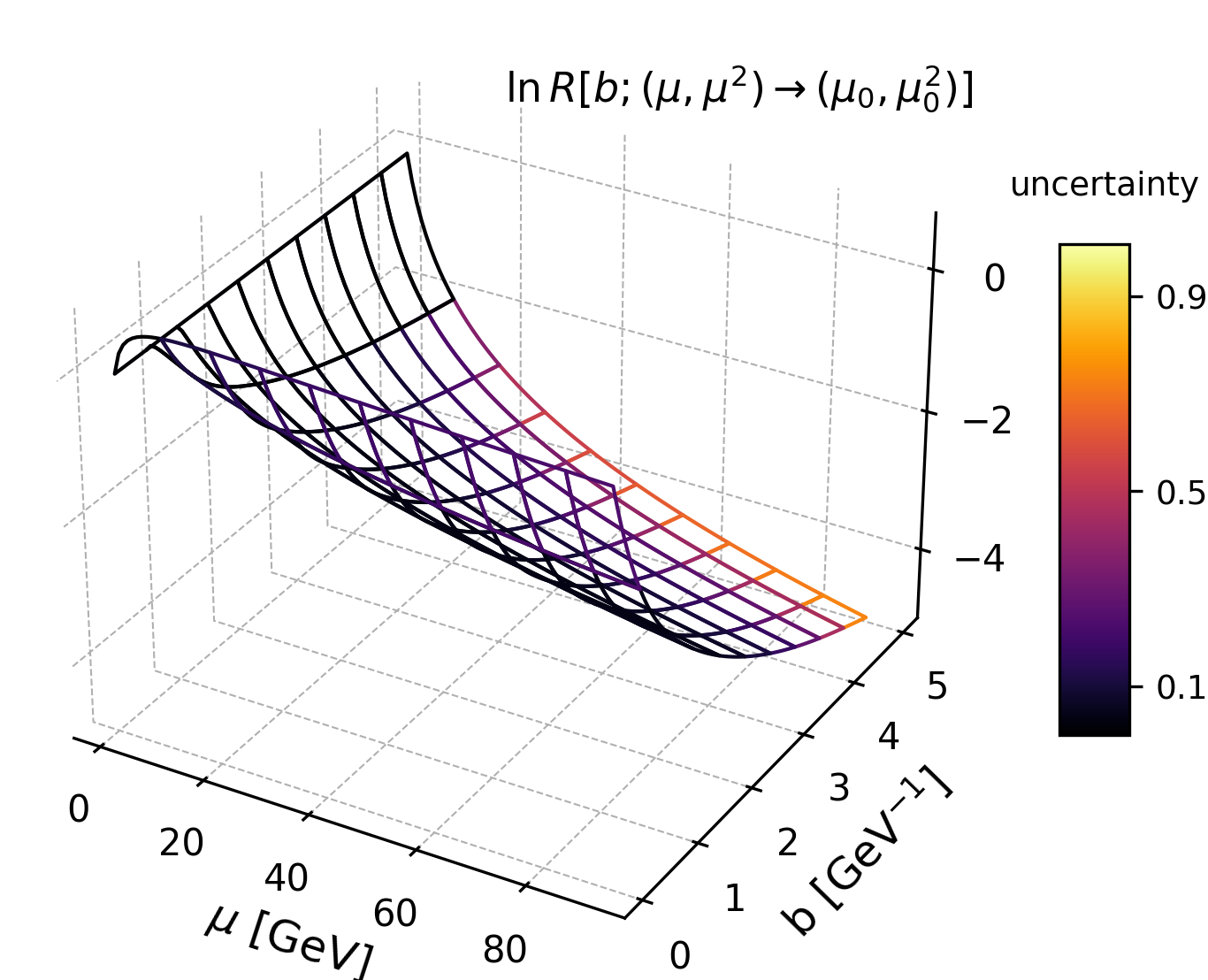}
\caption{\label{fig1} (top) CS kernel ${\cal D}(\mu,b)$ determined from DY transverse momentum spectra simulated with the PB method as proposed in~\cite{BermudezMartinez:2022ctj}, and (bottom) logarithm of the evolution factor $R [b; (\mu, \mu^2) \rightarrow (\mu_0, \mu_0^2)]$. The color scale indicates the corresponding absolute uncertainty.}
\end{figure}

The last step for obtaining the PB-TMD in the CSS framework at any given set of scales $(\mu, \zeta)$ is to evolve the Hankel transform of eq.~\ref{eq3} evaluated at $\mu_\text{OPE}$, using eq.~\ref{eq6}. The path independent expresion for the evolution factor $R$ can be written as~\cite{Scimemi:2018xaf}:
\begin{align}
    & R [b; (\mu, \zeta) \rightarrow (\mu_0, \zeta_0)] = \nonumber \\
    & \exp \Bigg\{ - \int_{\mu_0}^{\mu} \frac{d\mu'}{\mu'} (2 {\cal D} (\mu',b) + \gamma_V(\mu') )  \nonumber \nonumber \\ 
    &+ {\cal D}(\mu,b) \ln \left( \frac{\mu^2}{\zeta} \right) - {\cal D}(\mu_0,b) \ln \left( \frac{\mu_0^2}{\zeta_0} \right) \Bigg\},
\label{eq10}
\end{align}
where $\gamma_V$ is the anomalous dimension from the TMD vector form factor~\cite{Scimemi:2019cmh}. In order to evaluate the evolution factor $R$ for the case of PB, the corresponding CS kernel ${\cal D}(\mu,b)$ needs to be known. The CS kernel can be determined at the cross-section level from DY transverse momentum spectra, as has been demonstrated in~\cite{BermudezMartinez:2022ctj}. Its determination, using DY events simulated with the CASCADE event generator~\cite{CASCADE:2021bxe} via the PB approach is shown in Fig.~\ref{fig1} (top), for the case in which the TMD distribution PBset2~\cite{BermudezMartinez:2018fsv} is employed. I use DY production in $pp$ collisions for DY masses $Q=$ 12, 16, 20, 24 GeV, and center-of-mass energies $\sqrt{s} = $ 655.2, 873.6, 1092.0, 1310.0 GeV respectively. These values for $Q$ and $\sqrt{s}$ follow the choices in~\cite{BermudezMartinez:2022ctj} and ensure that the same ranges in longitudinal momentum fraction are probed for a maximum DY rapidity of 4. I use a bin size in transverse momentum of 0.05 GeV which allows to reach transverse distance values $b\sim 4-5$ GeV$^{-1}$. 

The systematic uncertainty in the CS determination is shown in Fig.~\ref{fig1} (top) in a color scale. The uncertainty includes the propagation of statistical uncertainty of the DY spectra simulations using the bootstrap method, from momentum space to position space. It also includes the uncertainty due to the finite bin size by varying the central value within each bin of the DY transverse momentum spectra.

In order to evaluate the evolution factor defined in eq.~\ref{eq10} the scale $\mu_0$ is set to $\sqrt{1.4~\text{GeV}^2}$ which is the value of the reference scale for the case of the PBset2 parton distribution. In addition $\zeta_0$ and $\zeta$ are set to $\mu_0^2$ and $\mu^2$ respectively, which allow to eliminate the logarithm terms of the exponent in eq.~\ref{eq10}. The resulting evolution factor $R [b; (\mu, \mu^2) \rightarrow (\mu_0, \mu_0^2)]$ is shown in Fig.~\ref{fig1} (bottom), where the color scale represents the uncertainty stemming from the propagation of systematic uncertainty on the CS kernel determination.

\begin{figure}[h]
\includegraphics[width=0.495\textwidth]{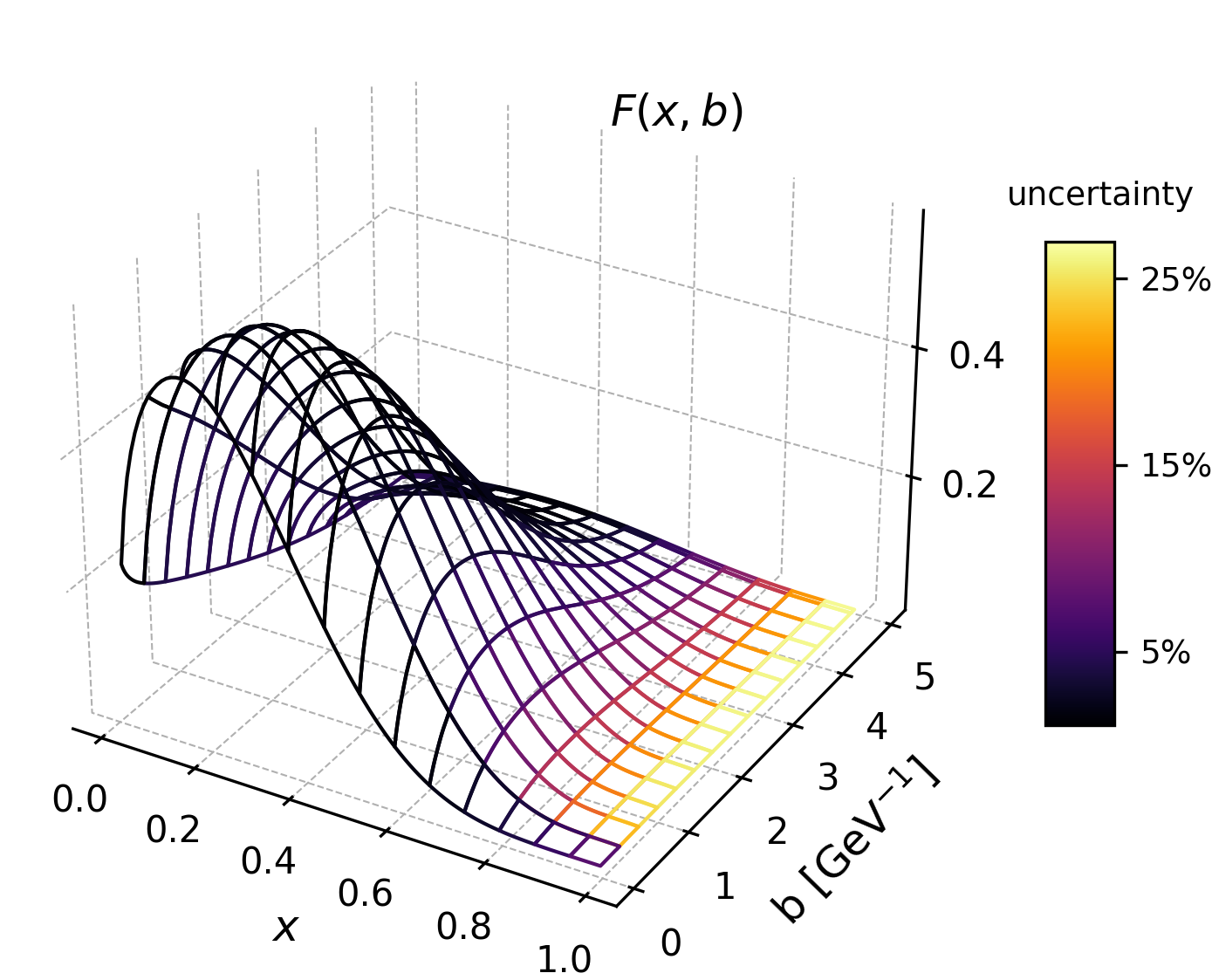}
\includegraphics[width=0.495\textwidth]{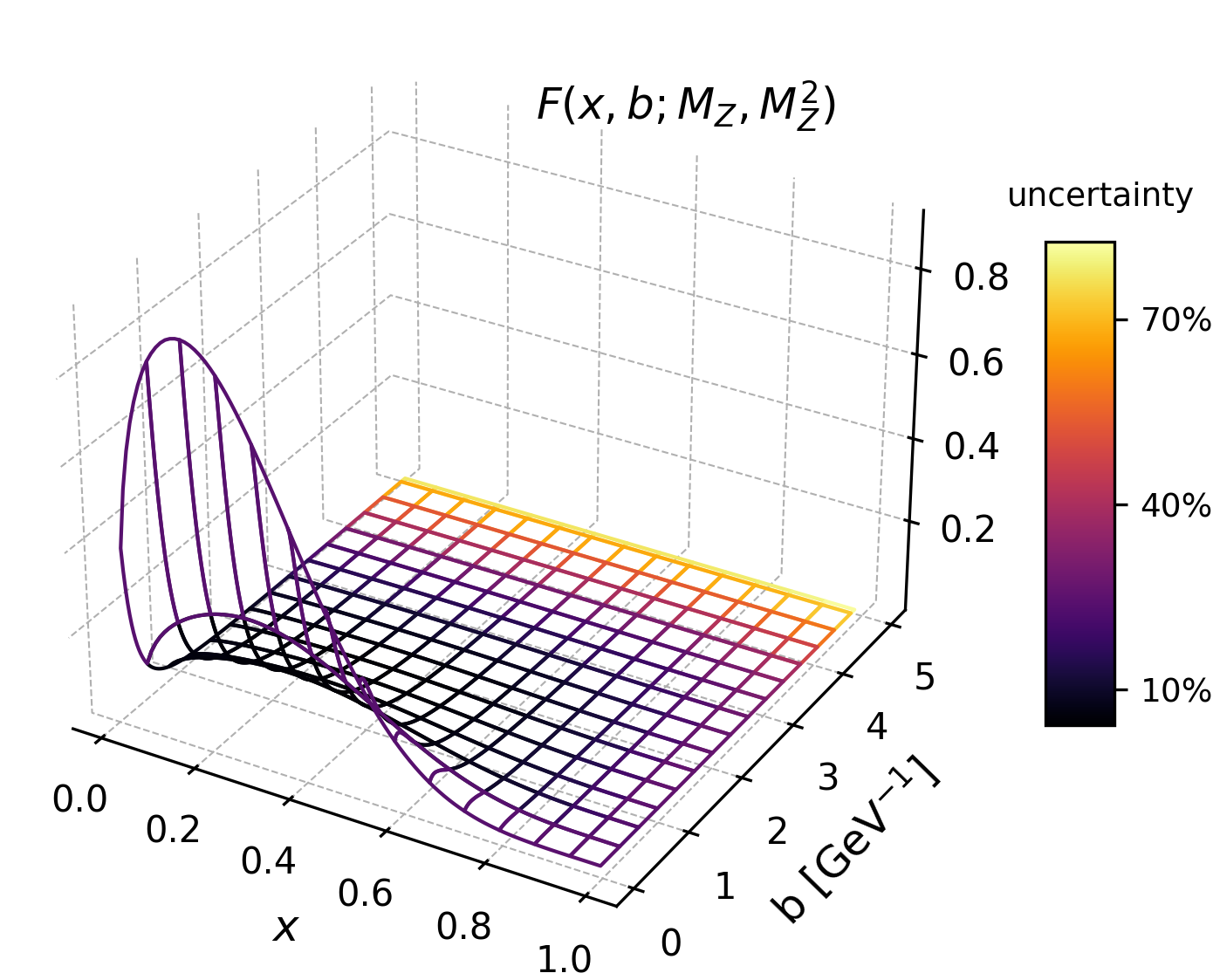}
\caption{\label{fig2} (top) up-valence quark starting TMD distribution, and (bottom) TMD evaluated at $\mu=M_Z$. The color scale indicates the corresponding relative uncertainty.}
\end{figure}

Following eq.~\ref{eq9}, the up-valence quark starting distribution $F(x,b)$ for the case of PBset2 can be written as:
\begin{align}
& F(x, b) = f (x, \frac{2e^{\gamma_E}}{b} + \mu_0) \nonumber \\
& \times \int_0^\infty dk_\perp k_\perp J_0(k_\perp b) g_{NP}(k_\perp^2) \;,
\label{eq11}
\end{align}
where the scale of the integrated collinear distribution is set to $\mu_\text{OPE}$ as defined in ep.~\ref{eq8}, and the integral corresponds to the inverse Hankel transform of the intrinsic transverse momentum distribution $g_{NP}$, which is a Gaussian distribution for the case of PBset2 ~\cite{BermudezMartinez:2018fsv}. The resulting distribution $F(x,b)$ is depicted in Fig.~\ref{fig2} (top), for the case of PBset2. The resulting uncertainty from the integrated PBset2 parametrization is available in TMDlib~\cite{Abdulov:2021ivr}, and is depicted with a color scale. 
The TMD distribution evaluated at the Z boson mass  $F(x, b; M_Z, M_Z)$ is obtained by evolving $F(x,b)$ with the factor $R [b; (M_Z, M_Z^2) \rightarrow (\mu_0, \mu_0^2)]$ resulting in the  expression:
\begin{align}
& F(x, b; M_Z, M_Z) = F(x, b) \nonumber \\ 
&\times \exp \Bigg\{ - \int_{\mu_0}^{M_Z} \frac{d\mu'}{\mu'} (2 {\cal D} (\mu',b) + \gamma_V(\mu') ) \Bigg\} \;,
\label{eq12}
\end{align}
The resulting $F(x, b; M_Z, M_Z)$ is shown in Fig.~\ref{fig2} (bottom). The uncertainty represented by the color scale includes the integrated TMD parametrization, as well as the propagation of the CS kernel systematic uncertainty. While at low scales the main uncertainty corresponds to the parametrization of the integrated TMD, at high scales the systematic uncertainty of the CS kernel propagated to the evolution factor in eq.~\ref{eq9} becomes dominant, especially at large $b$. 

\begin{figure}[h]
\includegraphics[width=0.49\textwidth]{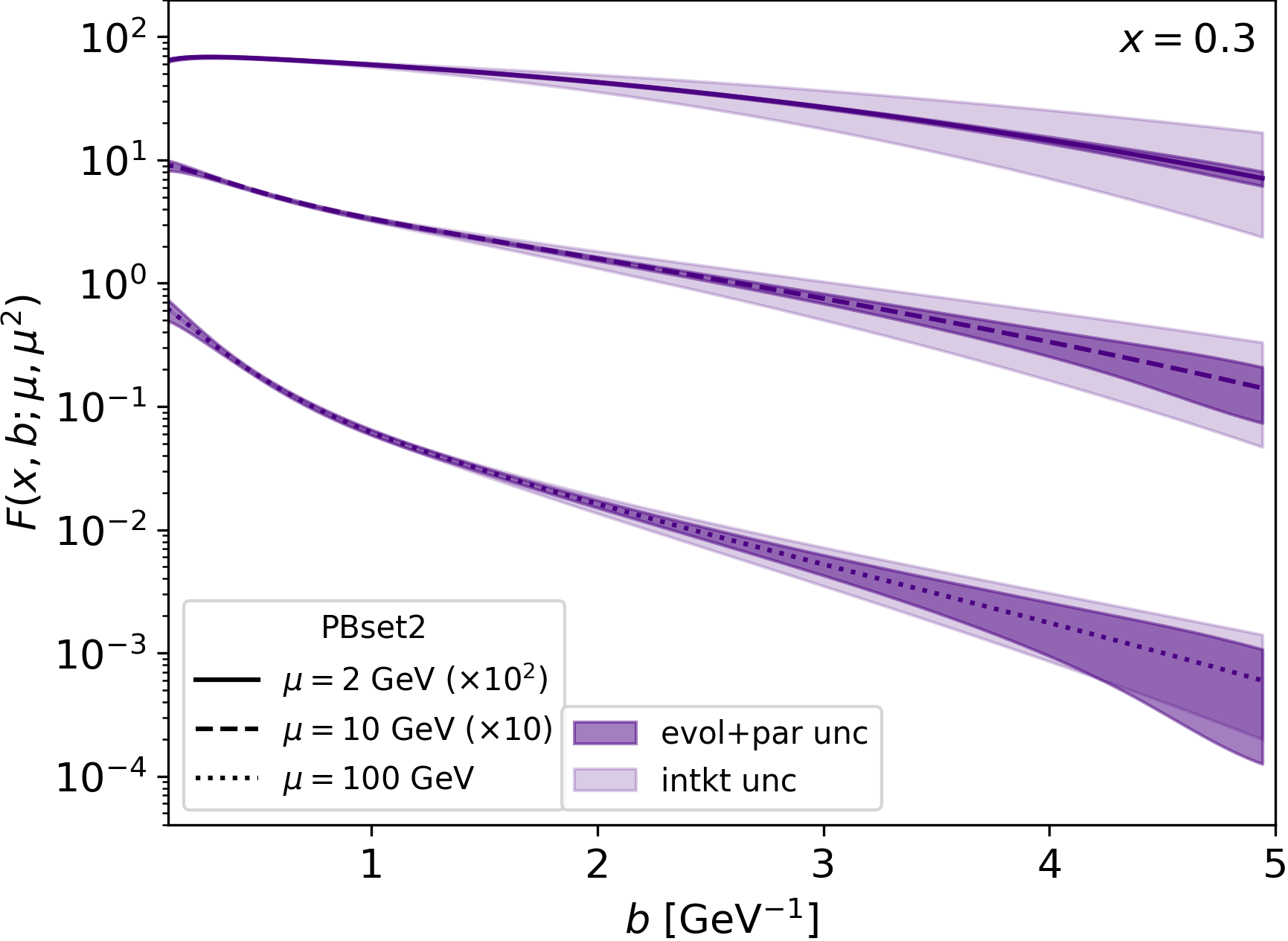}\\~\\
\includegraphics[width=0.49\textwidth]{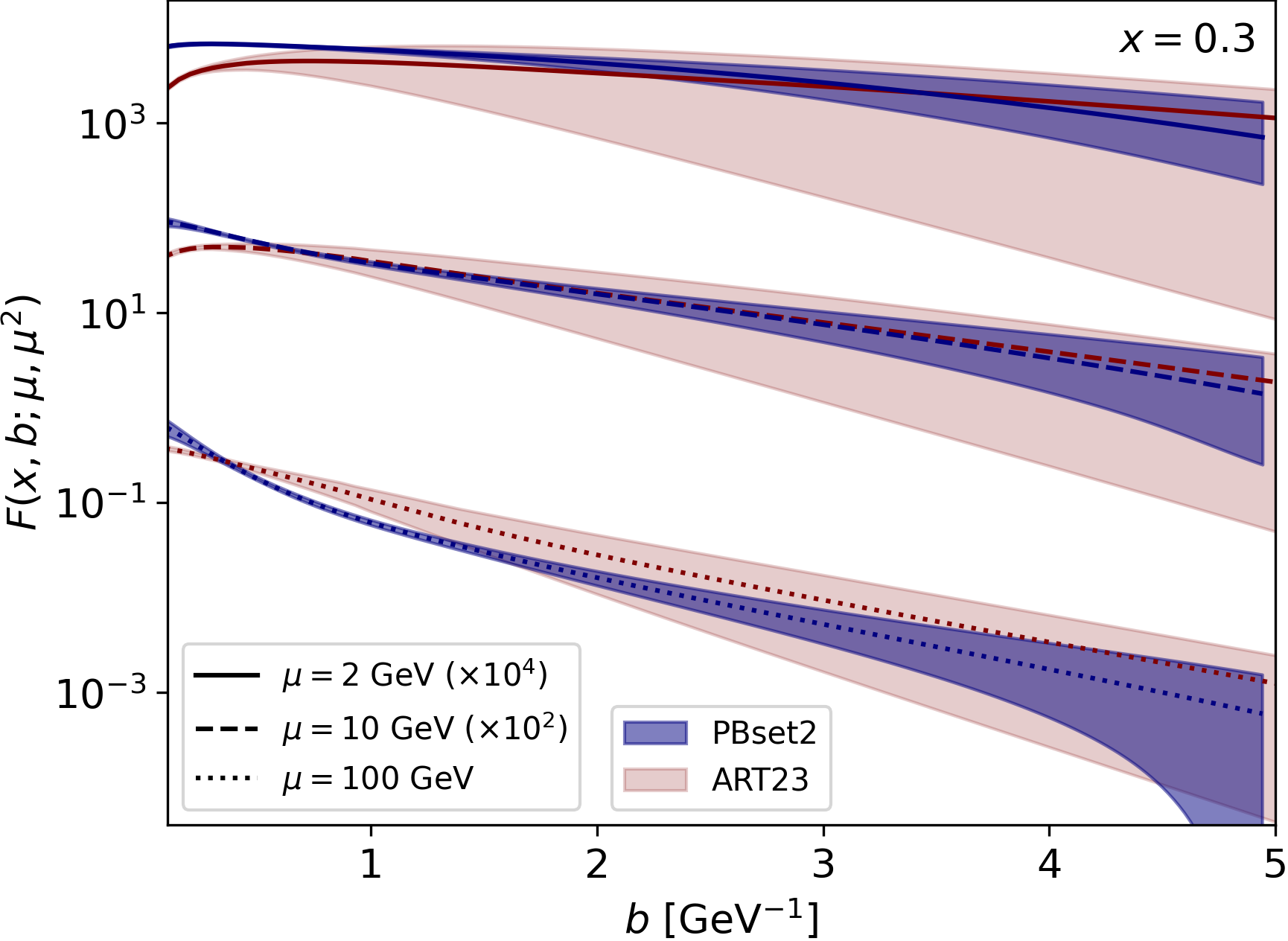}
\caption{\label{fig3} up-valence quark TMD at $x=0.3$, evaluated at the scales $\mu=2, 10, 100$ GeV. (top) PBset2 TMD distribution is shown. The dark violet band includes the propagation of the CS kernel systematic uncertainty and integrated TMD parametrization uncertainty, while the light violet band includes the variation of the width of the Gaussian intrinsic transverse momentum distribution. Distributions are scaled for better comparison. (bottom) PBset2 TMD distribution is compared to the ART23 TMD set~\cite{Moos:2023yfa}. The bands correspond to the respective total uncertainties. Distributions are scaled for better comparison}
\end{figure}

As depicted in Fig.~\ref{fig2}, when the scale of the process increases the TMD becomes narrower as a function of $b$. This can be better observed in Fig.~\ref{fig3} (top) where the TMD at fixed $x=0,3$, and as a function of $b$ is shown, evaluated at the scales $\mu=2, 10, 100$ GeV. This implies that the corresponding TMD in momentum space will have a stronger tail as a result of the evolution to higher scales. The uncertainty corresponding to the propagation of the CS kernel systematic uncertainty and integrated TMD parametrization is represented as a dark violet band in Fig.~\ref{fig3} (top). As the impact of the evolution factor in eq.~\ref{eq10} increases with increasing scale, the resulting uncertainty on the evolved TMD distribution also increases. In addition, Fig.~\ref{fig3} (top) also shows the uncertainty corresponding to the variation of the width of the Gaussian intrinsic transverse momentum distribution of PBset2 as a light violet band. The Gaussian width was varied between 0.3 and 0.5 GeV, based on the analysis of low transverse momentum DY data, which was performed in~\cite{BermudezMartinez:2020tys} using PBset2. At low scales, where the effect of evolution is small, the dominant uncertainty corresponds to the parametrization of the intrinsic transverse momentum distribution.

In Fig.~\ref{fig3} (bottom) the recently published ART23 TMD distribution~\cite{Moos:2023yfa} evaluated at $\mu=2, 10, 100$ GeV and shown in red color, is compared to PBset2 shown in blue color. The uncertainty band corresponding to the PBset2 distribution includes the propagation of the CS kernel systematic uncertainty, the integrated TMD parametrization, and the intrinsic transverse momentum Gaussian width variation. For the case of ART23 the band represents the envelope of its 1000 replicas and corresponds to the total uncertainty which includes experimental uncertainties and collinear PDF bias~\cite{Moos:2023yfa}. It is worth pointing out that the reason why the uncertainty on PBset2 is significantly lower than that corresponding to ART23 is that PBset2 does not suffer from collinear PDF bias. This is due to the fact that the collinear PDF corresponds in this case to the integrated PBset2. As discussed in~\cite{Moos:2023yfa} the uncertainty band for ART23 would be an order of magnitude smaller if PDF bias was to be neglected.
At low $b$ one can observe a significant difference between the TMDs, which is explained by the different order of the matching coefficients used in eq.~\ref{eq7}, which for the case of ART23 corresponds to N$^3$LO. At $\mu = 2$ GeV the choice of the starting distribution plays a role, in the case of ART23 the parametrized non-perturbative function in eq.~\ref{eq7} is dependent not only on $b$ but also on $x$. Due to the evolution, the effect of the different starting distributions decreases at larger scales and a better agreement between the TMDs is observed.  

{\bf Conclusions.} I have performed the first transformation of PB-TMDs to the CSS framework. I used the underlying CS kernel determined from simulated DY transverse momentum spectra to perform the evolution of the TMDs in $b$-space. The results include the effect of uncertainties from the parametrization of the collinear integrated TMDs and from the propagation of the systematic uncertainties arising from the CS kernel determination. I have shown the first comparison of TMDs obtained from the different frameworks, PBset2 and ART23. The results open the door for the the usage of PB-TMDs in CSS calculations, and also for the simulation of fully exclusive collision events using CSS TMDs within the PB Monte Carlo framework. This work also allows to look in more detail the systematic effects of collinear distributions in TMD fits, given that in the PB framework TMDs are obtained through fits of the integrated distributions, while in the CSS formalism the non-perturbative $b$-dependent part is fitted instead.

{\bf Acknowledgments.}
The author is thankful to David D'Enterria, Hannes Jung, Francesco Hautmann, Alexey Vladimirov for stimulating discussions, and additionally to Alexey Vladimirov for providing the grids corresponding to the ART23 TMD distribution.

\bibliography{biblio}

\begin{thebibliography}{38}%
\makeatletter
\providecommand \@ifxundefined [1]{%
 \@ifx{#1\undefined}
}%
\providecommand \@ifnum [1]{%
 \ifnum #1\expandafter \@firstoftwo
 \else \expandafter \@secondoftwo
 \fi
}%
\providecommand \@ifx [1]{%
 \ifx #1\expandafter \@firstoftwo
 \else \expandafter \@secondoftwo
 \fi
}%
\providecommand \natexlab [1]{#1}%
\providecommand \enquote  [1]{``#1''}%
\providecommand \bibnamefont  [1]{#1}%
\providecommand \bibfnamefont [1]{#1}%
\providecommand \citenamefont [1]{#1}%
\providecommand \href@noop [0]{\@secondoftwo}%
\providecommand \href [0]{\begingroup \@sanitize@url \@href}%
\providecommand \@href[1]{\@@startlink{#1}\@@href}%
\providecommand \@@href[1]{\endgroup#1\@@endlink}%
\providecommand \@sanitize@url [0]{\catcode `\\12\catcode `\$12\catcode
  `\&12\catcode `\#12\catcode `\^12\catcode `\_12\catcode `\%12\relax}%
\providecommand \@@startlink[1]{}%
\providecommand \@@endlink[0]{}%
\providecommand \url  [0]{\begingroup\@sanitize@url \@url }%
\providecommand \@url [1]{\endgroup\@href {#1}{\urlprefix }}%
\providecommand \urlprefix  [0]{URL }%
\providecommand \Eprint [0]{\href }%
\providecommand \doibase [0]{http://dx.doi.org/}%
\providecommand \selectlanguage [0]{\@gobble}%
\providecommand \bibinfo  [0]{\@secondoftwo}%
\providecommand \bibfield  [0]{\@secondoftwo}%
\providecommand \translation [1]{[#1]}%
\providecommand \BibitemOpen [0]{}%
\providecommand \bibitemStop [0]{}%
\providecommand \bibitemNoStop [0]{.\EOS\space}%
\providecommand \EOS [0]{\spacefactor3000\relax}%
\providecommand \BibitemShut  [1]{\csname bibitem#1\endcsname}%
\let\auto@bib@innerbib\@empty
\bibitem [{\citenamefont {Collins}\ \emph {et~al.}(1989)\citenamefont
  {Collins}, \citenamefont {Soper},\ and\ \citenamefont
  {Sterman}}]{Collins:1989gx}%
  \BibitemOpen
  \bibfield  {author} {\bibinfo {author} {\bibfnamefont {J.~C.}\ \bibnamefont
  {Collins}}, \bibinfo {author} {\bibfnamefont {D.~E.}\ \bibnamefont {Soper}},
  \ and\ \bibinfo {author} {\bibfnamefont {G.~F.}\ \bibnamefont {Sterman}},\
  }\href {\doibase 10.1142/9789814503266\textunderscore0001} {\bibfield
  {journal} {\bibinfo  {journal} {Adv. Ser. Direct. High Energy Phys.}\
  }\textbf {\bibinfo {volume} {5}},\ \bibinfo {pages} {1} (\bibinfo {year}
  {1989})},\ \Eprint {http://arxiv.org/abs/hep-ph/0409313}
  {arXiv:hep-ph/0409313} \BibitemShut {NoStop}%
\bibitem [{\citenamefont {Collins}(2013)}]{Collins:2011zzd}%
  \BibitemOpen
  \bibfield  {author} {\bibinfo {author} {\bibfnamefont {J.}~\bibnamefont
  {Collins}},\ }\href@noop {} {\emph {\bibinfo {title} {{Foundations of
  perturbative QCD}}}},\ Vol.~\bibinfo {volume} {32}\ (\bibinfo  {publisher}
  {Cambridge University Press},\ \bibinfo {year} {2013})\BibitemShut {NoStop}%
\bibitem [{\citenamefont {Echevarria}\ \emph {et~al.}(2012)\citenamefont
  {Echevarria}, \citenamefont {Idilbi},\ and\ \citenamefont
  {Scimemi}}]{Echevarria:2011epo}%
  \BibitemOpen
  \bibfield  {author} {\bibinfo {author} {\bibfnamefont {M.~G.}\ \bibnamefont
  {Echevarria}}, \bibinfo {author} {\bibfnamefont {A.}~\bibnamefont {Idilbi}},
  \ and\ \bibinfo {author} {\bibfnamefont {I.}~\bibnamefont {Scimemi}},\ }\href
  {\doibase 10.1007/JHEP07(2012)002} {\bibfield  {journal} {\bibinfo  {journal}
  {JHEP}\ }\textbf {\bibinfo {volume} {07}},\ \bibinfo {pages} {002} (\bibinfo
  {year} {2012})},\ \Eprint {http://arxiv.org/abs/1111.4996} {arXiv:1111.4996
  [hep-ph]} \BibitemShut {NoStop}%
\bibitem [{\citenamefont {Becher}\ and\ \citenamefont
  {Neubert}(2011)}]{Becher:2010tm}%
  \BibitemOpen
  \bibfield  {author} {\bibinfo {author} {\bibfnamefont {T.}~\bibnamefont
  {Becher}}\ and\ \bibinfo {author} {\bibfnamefont {M.}~\bibnamefont
  {Neubert}},\ }\href {\doibase 10.1140/epjc/s10052-011-1665-7} {\bibfield
  {journal} {\bibinfo  {journal} {Eur. Phys. J. C}\ }\textbf {\bibinfo {volume}
  {71}},\ \bibinfo {pages} {1665} (\bibinfo {year} {2011})},\ \Eprint
  {http://arxiv.org/abs/1007.4005} {arXiv:1007.4005 [hep-ph]} \BibitemShut
  {NoStop}%
\bibitem [{\citenamefont {Hautmann}\ \emph {et~al.}(2018)\citenamefont
  {Hautmann}, \citenamefont {Jung}, \citenamefont {Lelek}, \citenamefont
  {Radescu},\ and\ \citenamefont {Zlebcik}}]{Hautmann:2017fcj}%
  \BibitemOpen
  \bibfield  {author} {\bibinfo {author} {\bibfnamefont {F.}~\bibnamefont
  {Hautmann}}, \bibinfo {author} {\bibfnamefont {H.}~\bibnamefont {Jung}},
  \bibinfo {author} {\bibfnamefont {A.}~\bibnamefont {Lelek}}, \bibinfo
  {author} {\bibfnamefont {V.}~\bibnamefont {Radescu}}, \ and\ \bibinfo
  {author} {\bibfnamefont {R.}~\bibnamefont {Zlebcik}},\ }\href {\doibase
  10.1007/JHEP01(2018)070} {\bibfield  {journal} {\bibinfo  {journal} {JHEP}\
  }\textbf {\bibinfo {volume} {01}},\ \bibinfo {pages} {070} (\bibinfo {year}
  {2018})},\ \Eprint {http://arxiv.org/abs/1708.03279} {arXiv:1708.03279
  [hep-ph]} \BibitemShut {NoStop}%
\bibitem [{\citenamefont {Hautmann}\ \emph {et~al.}(2017)\citenamefont
  {Hautmann}, \citenamefont {Jung}, \citenamefont {Lelek}, \citenamefont
  {Radescu},\ and\ \citenamefont {Zlebcik}}]{Hautmann:2017xtx}%
  \BibitemOpen
  \bibfield  {author} {\bibinfo {author} {\bibfnamefont {F.}~\bibnamefont
  {Hautmann}}, \bibinfo {author} {\bibfnamefont {H.}~\bibnamefont {Jung}},
  \bibinfo {author} {\bibfnamefont {A.}~\bibnamefont {Lelek}}, \bibinfo
  {author} {\bibfnamefont {V.}~\bibnamefont {Radescu}}, \ and\ \bibinfo
  {author} {\bibfnamefont {R.}~\bibnamefont {Zlebcik}},\ }\href {\doibase
  10.1016/j.physletb.2017.07.005} {\bibfield  {journal} {\bibinfo  {journal}
  {Phys. Lett. B}\ }\textbf {\bibinfo {volume} {772}},\ \bibinfo {pages} {446}
  (\bibinfo {year} {2017})},\ \Eprint {http://arxiv.org/abs/1704.01757}
  {arXiv:1704.01757 [hep-ph]} \BibitemShut {NoStop}%
\bibitem [{\citenamefont {Bermudez~Martinez}\ and\ \citenamefont
  {Vladimirov}(2022)}]{BermudezMartinez:2022ctj}%
  \BibitemOpen
  \bibfield  {author} {\bibinfo {author} {\bibfnamefont {A.}~\bibnamefont
  {Bermudez~Martinez}}\ and\ \bibinfo {author} {\bibfnamefont {A.}~\bibnamefont
  {Vladimirov}},\ }\href {\doibase 10.1103/PhysRevD.106.L091501} {\bibfield
  {journal} {\bibinfo  {journal} {Phys. Rev. D}\ }\textbf {\bibinfo {volume}
  {106}},\ \bibinfo {pages} {L091501} (\bibinfo {year} {2022})},\ \Eprint
  {http://arxiv.org/abs/2206.01105} {arXiv:2206.01105 [hep-ph]} \BibitemShut
  {NoStop}%
\bibitem [{\citenamefont {Gribov}\ and\ \citenamefont
  {Lipatov}(1972)}]{Gribov:1972ri}%
  \BibitemOpen
  \bibfield  {author} {\bibinfo {author} {\bibfnamefont {V.~N.}\ \bibnamefont
  {Gribov}}\ and\ \bibinfo {author} {\bibfnamefont {L.~N.}\ \bibnamefont
  {Lipatov}},\ }\href@noop {} {\bibfield  {journal} {\bibinfo  {journal} {Sov.
  J. Nucl. Phys.}\ }\textbf {\bibinfo {volume} {15}},\ \bibinfo {pages} {438}
  (\bibinfo {year} {1972})}\BibitemShut {NoStop}%
\bibitem [{\citenamefont {Lipatov}(1974)}]{Lipatov:1974qm}%
  \BibitemOpen
  \bibfield  {author} {\bibinfo {author} {\bibfnamefont {L.~N.}\ \bibnamefont
  {Lipatov}},\ }\href@noop {} {\bibfield  {journal} {\bibinfo  {journal} {Yad.
  Fiz.}\ }\textbf {\bibinfo {volume} {20}},\ \bibinfo {pages} {181} (\bibinfo
  {year} {1974})}\BibitemShut {NoStop}%
\bibitem [{\citenamefont {Altarelli}\ and\ \citenamefont
  {Parisi}(1977)}]{Altarelli:1977zs}%
  \BibitemOpen
  \bibfield  {author} {\bibinfo {author} {\bibfnamefont {G.}~\bibnamefont
  {Altarelli}}\ and\ \bibinfo {author} {\bibfnamefont {G.}~\bibnamefont
  {Parisi}},\ }\href {\doibase 10.1016/0550-3213(77)90384-4} {\bibfield
  {journal} {\bibinfo  {journal} {Nucl. Phys. B}\ }\textbf {\bibinfo {volume}
  {126}},\ \bibinfo {pages} {298} (\bibinfo {year} {1977})}\BibitemShut
  {NoStop}%
\bibitem [{\citenamefont {Dokshitzer}(1977)}]{Dokshitzer:1977sg}%
  \BibitemOpen
  \bibfield  {author} {\bibinfo {author} {\bibfnamefont {Y.~L.}\ \bibnamefont
  {Dokshitzer}},\ }\href@noop {} {\bibfield  {journal} {\bibinfo  {journal}
  {Sov. Phys. JETP}\ }\textbf {\bibinfo {volume} {46}},\ \bibinfo {pages} {641}
  (\bibinfo {year} {1977})}\BibitemShut {NoStop}%
\bibitem [{\citenamefont {Bermudez~Martinez}\ \emph
  {et~al.}(2019{\natexlab{a}})\citenamefont {Bermudez~Martinez} \emph
  {et~al.}}]{BermudezMartinez:2019anj}%
  \BibitemOpen
  \bibfield  {author} {\bibinfo {author} {\bibfnamefont {A.}~\bibnamefont
  {Bermudez~Martinez}} \emph {et~al.},\ }\href {\doibase
  10.1103/PhysRevD.100.074027} {\bibfield  {journal} {\bibinfo  {journal}
  {Phys. Rev. D}\ }\textbf {\bibinfo {volume} {100}},\ \bibinfo {pages}
  {074027} (\bibinfo {year} {2019}{\natexlab{a}})},\ \Eprint
  {http://arxiv.org/abs/1906.00919} {arXiv:1906.00919 [hep-ph]} \BibitemShut
  {NoStop}%
\bibitem [{\citenamefont {Bermudez~Martinez}\ \emph {et~al.}(2020)\citenamefont
  {Bermudez~Martinez} \emph {et~al.}}]{BermudezMartinez:2020tys}%
  \BibitemOpen
  \bibfield  {author} {\bibinfo {author} {\bibfnamefont {A.}~\bibnamefont
  {Bermudez~Martinez}} \emph {et~al.},\ }\href {\doibase
  10.1140/epjc/s10052-020-8136-y} {\bibfield  {journal} {\bibinfo  {journal}
  {Eur. Phys. J. C}\ }\textbf {\bibinfo {volume} {80}},\ \bibinfo {pages} {598}
  (\bibinfo {year} {2020})},\ \Eprint {http://arxiv.org/abs/2001.06488}
  {arXiv:2001.06488 [hep-ph]} \BibitemShut {NoStop}%
\bibitem [{\citenamefont {Martinez}\ \emph {et~al.}(2022)\citenamefont
  {Martinez}, \citenamefont {Estevez~Banos}, \citenamefont {Jung},
  \citenamefont {Lidrych}, \citenamefont {Mendizabal}, \citenamefont
  {Monfared}, \citenamefont {Wang},\ and\ \citenamefont
  {Yang}}]{Martinez:2021mzy}%
  \BibitemOpen
  \bibfield  {author} {\bibinfo {author} {\bibfnamefont {A.~B.}\ \bibnamefont
  {Martinez}}, \bibinfo {author} {\bibfnamefont {L.~I.}\ \bibnamefont
  {Estevez~Banos}}, \bibinfo {author} {\bibfnamefont {H.}~\bibnamefont {Jung}},
  \bibinfo {author} {\bibfnamefont {J.}~\bibnamefont {Lidrych}}, \bibinfo
  {author} {\bibfnamefont {M.}~\bibnamefont {Mendizabal}}, \bibinfo {author}
  {\bibfnamefont {S.~T.}\ \bibnamefont {Monfared}}, \bibinfo {author}
  {\bibfnamefont {Q.}~\bibnamefont {Wang}}, \ and\ \bibinfo {author}
  {\bibfnamefont {H.}~\bibnamefont {Yang}},\ }\href {\doibase
  10.22323/1.398.0386} {\bibfield  {journal} {\bibinfo  {journal} {PoS}\
  }\textbf {\bibinfo {volume} {EPS-HEP2021}},\ \bibinfo {pages} {386} (\bibinfo
  {year} {2022})},\ \Eprint {http://arxiv.org/abs/2111.03582} {arXiv:2111.03582
  [hep-ph]} \BibitemShut {NoStop}%
\bibitem [{\citenamefont {Sirunyan}\ \emph {et~al.}(2019)\citenamefont
  {Sirunyan} \emph {et~al.}}]{CMS:2019raw}%
  \BibitemOpen
  \bibfield  {author} {\bibinfo {author} {\bibfnamefont {A.~M.}\ \bibnamefont
  {Sirunyan}} \emph {et~al.} (\bibinfo {collaboration} {CMS}),\ }\href
  {\doibase 10.1007/JHEP12(2019)061} {\bibfield  {journal} {\bibinfo  {journal}
  {JHEP}\ }\textbf {\bibinfo {volume} {12}},\ \bibinfo {pages} {061} (\bibinfo
  {year} {2019})},\ \Eprint {http://arxiv.org/abs/1909.04133} {arXiv:1909.04133
  [hep-ex]} \BibitemShut {NoStop}%
\bibitem [{\citenamefont {Martinez}\ \emph {et~al.}(2021)\citenamefont
  {Martinez}, \citenamefont {Hautmann},\ and\ \citenamefont
  {Mangano}}]{Martinez:2021chk}%
  \BibitemOpen
  \bibfield  {author} {\bibinfo {author} {\bibfnamefont {A.~B.}\ \bibnamefont
  {Martinez}}, \bibinfo {author} {\bibfnamefont {F.}~\bibnamefont {Hautmann}},
  \ and\ \bibinfo {author} {\bibfnamefont {M.~L.}\ \bibnamefont {Mangano}},\
  }\href {\doibase 10.1016/j.physletb.2021.136700} {\bibfield  {journal}
  {\bibinfo  {journal} {Phys. Lett. B}\ }\textbf {\bibinfo {volume} {822}},\
  \bibinfo {pages} {136700} (\bibinfo {year} {2021})},\ \Eprint
  {http://arxiv.org/abs/2107.01224} {arXiv:2107.01224 [hep-ph]} \BibitemShut
  {NoStop}%
\bibitem [{\citenamefont {Bermudez~Martinez}\ \emph {et~al.}(2022)\citenamefont
  {Bermudez~Martinez}, \citenamefont {Hautmann},\ and\ \citenamefont
  {Mangano}}]{BermudezMartinez:2022bpj}%
  \BibitemOpen
  \bibfield  {author} {\bibinfo {author} {\bibfnamefont {A.}~\bibnamefont
  {Bermudez~Martinez}}, \bibinfo {author} {\bibfnamefont {F.}~\bibnamefont
  {Hautmann}}, \ and\ \bibinfo {author} {\bibfnamefont {M.~L.}\ \bibnamefont
  {Mangano}},\ }\href {\doibase 10.1007/JHEP09(2022)060} {\bibfield  {journal}
  {\bibinfo  {journal} {JHEP}\ }\textbf {\bibinfo {volume} {09}},\ \bibinfo
  {pages} {060} (\bibinfo {year} {2022})},\ \Eprint
  {http://arxiv.org/abs/2208.02276} {arXiv:2208.02276 [hep-ph]} \BibitemShut
  {NoStop}%
\bibitem [{\citenamefont {Bermudez~Martinez}\ \emph
  {et~al.}(2019{\natexlab{b}})\citenamefont {Bermudez~Martinez}, \citenamefont
  {Connor}, \citenamefont {Jung}, \citenamefont {Lelek}, \citenamefont
  {\v{Z}leb\v{c}\'\i{}k}, \citenamefont {Hautmann},\ and\ \citenamefont
  {Radescu}}]{BermudezMartinez:2018fsv}%
  \BibitemOpen
  \bibfield  {author} {\bibinfo {author} {\bibfnamefont {A.}~\bibnamefont
  {Bermudez~Martinez}}, \bibinfo {author} {\bibfnamefont {P.}~\bibnamefont
  {Connor}}, \bibinfo {author} {\bibfnamefont {H.}~\bibnamefont {Jung}},
  \bibinfo {author} {\bibfnamefont {A.}~\bibnamefont {Lelek}}, \bibinfo
  {author} {\bibfnamefont {R.}~\bibnamefont {\v{Z}leb\v{c}\'\i{}k}}, \bibinfo
  {author} {\bibfnamefont {F.}~\bibnamefont {Hautmann}}, \ and\ \bibinfo
  {author} {\bibfnamefont {V.}~\bibnamefont {Radescu}},\ }\href {\doibase
  10.1103/PhysRevD.99.074008} {\bibfield  {journal} {\bibinfo  {journal} {Phys.
  Rev. D}\ }\textbf {\bibinfo {volume} {99}},\ \bibinfo {pages} {074008}
  (\bibinfo {year} {2019}{\natexlab{b}})},\ \Eprint
  {http://arxiv.org/abs/1804.11152} {arXiv:1804.11152 [hep-ph]} \BibitemShut
  {NoStop}%
\bibitem [{\citenamefont {Abramowicz}\ \emph {et~al.}(2015)\citenamefont
  {Abramowicz} \emph {et~al.}}]{H1:2015ubc}%
  \BibitemOpen
  \bibfield  {author} {\bibinfo {author} {\bibfnamefont {H.}~\bibnamefont
  {Abramowicz}} \emph {et~al.} (\bibinfo {collaboration} {H1, ZEUS}),\ }\href
  {\doibase 10.1140/epjc/s10052-015-3710-4} {\bibfield  {journal} {\bibinfo
  {journal} {Eur. Phys. J. C}\ }\textbf {\bibinfo {volume} {75}},\ \bibinfo
  {pages} {580} (\bibinfo {year} {2015})},\ \Eprint
  {http://arxiv.org/abs/1506.06042} {arXiv:1506.06042 [hep-ex]} \BibitemShut
  {NoStop}%
\bibitem [{\citenamefont {Scimemi}\ and\ \citenamefont
  {Vladimirov}(2018{\natexlab{a}})}]{Scimemi:2018xaf}%
  \BibitemOpen
  \bibfield  {author} {\bibinfo {author} {\bibfnamefont {I.}~\bibnamefont
  {Scimemi}}\ and\ \bibinfo {author} {\bibfnamefont {A.}~\bibnamefont
  {Vladimirov}},\ }\href {\doibase 10.1007/JHEP08(2018)003} {\bibfield
  {journal} {\bibinfo  {journal} {JHEP}\ }\textbf {\bibinfo {volume} {08}},\
  \bibinfo {pages} {003} (\bibinfo {year} {2018}{\natexlab{a}})},\ \Eprint
  {http://arxiv.org/abs/1803.11089} {arXiv:1803.11089 [hep-ph]} \BibitemShut
  {NoStop}%
\bibitem [{\citenamefont {Scimemi}\ and\ \citenamefont
  {Vladimirov}(2020)}]{Scimemi:2019cmh}%
  \BibitemOpen
  \bibfield  {author} {\bibinfo {author} {\bibfnamefont {I.}~\bibnamefont
  {Scimemi}}\ and\ \bibinfo {author} {\bibfnamefont {A.}~\bibnamefont
  {Vladimirov}},\ }\href {\doibase 10.1007/JHEP06(2020)137} {\bibfield
  {journal} {\bibinfo  {journal} {JHEP}\ }\textbf {\bibinfo {volume} {06}},\
  \bibinfo {pages} {137} (\bibinfo {year} {2020})},\ \Eprint
  {http://arxiv.org/abs/1912.06532} {arXiv:1912.06532 [hep-ph]} \BibitemShut
  {NoStop}%
\bibitem [{\citenamefont {Scimemi}\ and\ \citenamefont
  {Vladimirov}(2018{\natexlab{b}})}]{Scimemi:2017etj}%
  \BibitemOpen
  \bibfield  {author} {\bibinfo {author} {\bibfnamefont {I.}~\bibnamefont
  {Scimemi}}\ and\ \bibinfo {author} {\bibfnamefont {A.}~\bibnamefont
  {Vladimirov}},\ }\href {\doibase 10.1140/epjc/s10052-018-5557-y} {\bibfield
  {journal} {\bibinfo  {journal} {Eur. Phys. J. C}\ }\textbf {\bibinfo {volume}
  {78}},\ \bibinfo {pages} {89} (\bibinfo {year} {2018}{\natexlab{b}})},\
  \Eprint {http://arxiv.org/abs/1706.01473} {arXiv:1706.01473 [hep-ph]}
  \BibitemShut {NoStop}%
\bibitem [{\citenamefont {Bacchetta}\ \emph {et~al.}(2020)\citenamefont
  {Bacchetta}, \citenamefont {Bertone}, \citenamefont {Bissolotti},
  \citenamefont {Bozzi}, \citenamefont {Delcarro}, \citenamefont {Piacenza},\
  and\ \citenamefont {Radici}}]{Bacchetta:2019sam}%
  \BibitemOpen
  \bibfield  {author} {\bibinfo {author} {\bibfnamefont {A.}~\bibnamefont
  {Bacchetta}}, \bibinfo {author} {\bibfnamefont {V.}~\bibnamefont {Bertone}},
  \bibinfo {author} {\bibfnamefont {C.}~\bibnamefont {Bissolotti}}, \bibinfo
  {author} {\bibfnamefont {G.}~\bibnamefont {Bozzi}}, \bibinfo {author}
  {\bibfnamefont {F.}~\bibnamefont {Delcarro}}, \bibinfo {author}
  {\bibfnamefont {F.}~\bibnamefont {Piacenza}}, \ and\ \bibinfo {author}
  {\bibfnamefont {M.}~\bibnamefont {Radici}},\ }\href {\doibase
  10.1007/JHEP07(2020)117} {\bibfield  {journal} {\bibinfo  {journal} {JHEP}\
  }\textbf {\bibinfo {volume} {07}},\ \bibinfo {pages} {117} (\bibinfo {year}
  {2020})},\ \Eprint {http://arxiv.org/abs/1912.07550} {arXiv:1912.07550
  [hep-ph]} \BibitemShut {NoStop}%
\bibitem [{\citenamefont {Bertone}\ \emph {et~al.}(2019)\citenamefont
  {Bertone}, \citenamefont {Scimemi},\ and\ \citenamefont
  {Vladimirov}}]{Bertone:2019nxa}%
  \BibitemOpen
  \bibfield  {author} {\bibinfo {author} {\bibfnamefont {V.}~\bibnamefont
  {Bertone}}, \bibinfo {author} {\bibfnamefont {I.}~\bibnamefont {Scimemi}}, \
  and\ \bibinfo {author} {\bibfnamefont {A.}~\bibnamefont {Vladimirov}},\
  }\href {\doibase 10.1007/JHEP06(2019)028} {\bibfield  {journal} {\bibinfo
  {journal} {JHEP}\ }\textbf {\bibinfo {volume} {06}},\ \bibinfo {pages} {028}
  (\bibinfo {year} {2019})},\ \Eprint {http://arxiv.org/abs/1902.08474}
  {arXiv:1902.08474 [hep-ph]} \BibitemShut {NoStop}%
\bibitem [{\citenamefont {Bacchetta}\ \emph {et~al.}(2022)\citenamefont
  {Bacchetta}, \citenamefont {Bertone}, \citenamefont {Bissolotti},
  \citenamefont {Bozzi}, \citenamefont {Cerutti}, \citenamefont {Piacenza},
  \citenamefont {Radici},\ and\ \citenamefont {Signori}}]{Bacchetta:2022awv}%
  \BibitemOpen
  \bibfield  {author} {\bibinfo {author} {\bibfnamefont {A.}~\bibnamefont
  {Bacchetta}}, \bibinfo {author} {\bibfnamefont {V.}~\bibnamefont {Bertone}},
  \bibinfo {author} {\bibfnamefont {C.}~\bibnamefont {Bissolotti}}, \bibinfo
  {author} {\bibfnamefont {G.}~\bibnamefont {Bozzi}}, \bibinfo {author}
  {\bibfnamefont {M.}~\bibnamefont {Cerutti}}, \bibinfo {author} {\bibfnamefont
  {F.}~\bibnamefont {Piacenza}}, \bibinfo {author} {\bibfnamefont
  {M.}~\bibnamefont {Radici}}, \ and\ \bibinfo {author} {\bibfnamefont
  {A.}~\bibnamefont {Signori}} (\bibinfo {collaboration} {MAP
  (Multi-dimensional Analyses of Partonic distributions)}),\ }\href {\doibase
  10.1007/JHEP10(2022)127} {\bibfield  {journal} {\bibinfo  {journal} {JHEP}\
  }\textbf {\bibinfo {volume} {10}},\ \bibinfo {pages} {127} (\bibinfo {year}
  {2022})},\ \Eprint {http://arxiv.org/abs/2206.07598} {arXiv:2206.07598
  [hep-ph]} \BibitemShut {NoStop}%
\bibitem [{\citenamefont {Moos}\ \emph {et~al.}(2023)\citenamefont {Moos},
  \citenamefont {Scimemi}, \citenamefont {Vladimirov},\ and\ \citenamefont
  {Zurita}}]{Moos:2023yfa}%
  \BibitemOpen
  \bibfield  {author} {\bibinfo {author} {\bibfnamefont {V.}~\bibnamefont
  {Moos}}, \bibinfo {author} {\bibfnamefont {I.}~\bibnamefont {Scimemi}},
  \bibinfo {author} {\bibfnamefont {A.}~\bibnamefont {Vladimirov}}, \ and\
  \bibinfo {author} {\bibfnamefont {P.}~\bibnamefont {Zurita}},\ }\href@noop {}
  {\  (\bibinfo {year} {2023})},\ \Eprint {http://arxiv.org/abs/2305.07473}
  {arXiv:2305.07473 [hep-ph]} \BibitemShut {NoStop}%
\bibitem [{\citenamefont {Ball}\ \emph {et~al.}(2017)\citenamefont {Ball} \emph
  {et~al.}}]{NNPDF:2017mvq}%
  \BibitemOpen
  \bibfield  {author} {\bibinfo {author} {\bibfnamefont {R.~D.}\ \bibnamefont
  {Ball}} \emph {et~al.} (\bibinfo {collaboration} {NNPDF}),\ }\href {\doibase
  10.1140/epjc/s10052-017-5199-5} {\bibfield  {journal} {\bibinfo  {journal}
  {Eur. Phys. J. C}\ }\textbf {\bibinfo {volume} {77}},\ \bibinfo {pages} {663}
  (\bibinfo {year} {2017})},\ \Eprint {http://arxiv.org/abs/1706.00428}
  {arXiv:1706.00428 [hep-ph]} \BibitemShut {NoStop}%
\bibitem [{\citenamefont {Hou}\ \emph {et~al.}(2021)\citenamefont {Hou} \emph
  {et~al.}}]{Hou:2019efy}%
  \BibitemOpen
  \bibfield  {author} {\bibinfo {author} {\bibfnamefont {T.-J.}\ \bibnamefont
  {Hou}} \emph {et~al.},\ }\href {\doibase 10.1103/PhysRevD.103.014013}
  {\bibfield  {journal} {\bibinfo  {journal} {Phys. Rev. D}\ }\textbf {\bibinfo
  {volume} {103}},\ \bibinfo {pages} {014013} (\bibinfo {year} {2021})},\
  \Eprint {http://arxiv.org/abs/1912.10053} {arXiv:1912.10053 [hep-ph]}
  \BibitemShut {NoStop}%
\bibitem [{\citenamefont {Bailey}\ \emph {et~al.}(2021)\citenamefont {Bailey},
  \citenamefont {Cridge}, \citenamefont {Harland-Lang}, \citenamefont
  {Martin},\ and\ \citenamefont {Thorne}}]{Bailey:2020ooq}%
  \BibitemOpen
  \bibfield  {author} {\bibinfo {author} {\bibfnamefont {S.}~\bibnamefont
  {Bailey}}, \bibinfo {author} {\bibfnamefont {T.}~\bibnamefont {Cridge}},
  \bibinfo {author} {\bibfnamefont {L.~A.}\ \bibnamefont {Harland-Lang}},
  \bibinfo {author} {\bibfnamefont {A.~D.}\ \bibnamefont {Martin}}, \ and\
  \bibinfo {author} {\bibfnamefont {R.~S.}\ \bibnamefont {Thorne}},\ }\href
  {\doibase 10.1140/epjc/s10052-021-09057-0} {\bibfield  {journal} {\bibinfo
  {journal} {Eur. Phys. J. C}\ }\textbf {\bibinfo {volume} {81}},\ \bibinfo
  {pages} {341} (\bibinfo {year} {2021})},\ \Eprint
  {http://arxiv.org/abs/2012.04684} {arXiv:2012.04684 [hep-ph]} \BibitemShut
  {NoStop}%
\bibitem [{\citenamefont {Bury}\ \emph {et~al.}(2022)\citenamefont {Bury},
  \citenamefont {Hautmann}, \citenamefont {Leal-Gomez}, \citenamefont
  {Scimemi}, \citenamefont {Vladimirov},\ and\ \citenamefont
  {Zurita}}]{Bury:2022czx}%
  \BibitemOpen
  \bibfield  {author} {\bibinfo {author} {\bibfnamefont {M.}~\bibnamefont
  {Bury}}, \bibinfo {author} {\bibfnamefont {F.}~\bibnamefont {Hautmann}},
  \bibinfo {author} {\bibfnamefont {S.}~\bibnamefont {Leal-Gomez}}, \bibinfo
  {author} {\bibfnamefont {I.}~\bibnamefont {Scimemi}}, \bibinfo {author}
  {\bibfnamefont {A.}~\bibnamefont {Vladimirov}}, \ and\ \bibinfo {author}
  {\bibfnamefont {P.}~\bibnamefont {Zurita}},\ }\href {\doibase
  10.1007/JHEP10(2022)118} {\bibfield  {journal} {\bibinfo  {journal} {JHEP}\
  }\textbf {\bibinfo {volume} {10}},\ \bibinfo {pages} {118} (\bibinfo {year}
  {2022})},\ \Eprint {http://arxiv.org/abs/2201.07114} {arXiv:2201.07114
  [hep-ph]} \BibitemShut {NoStop}%
\bibitem [{\citenamefont {Vladimirov}(2020)}]{Vladimirov:2020umg}%
  \BibitemOpen
  \bibfield  {author} {\bibinfo {author} {\bibfnamefont {A.~A.}\ \bibnamefont
  {Vladimirov}},\ }\href {\doibase 10.1103/PhysRevLett.125.192002} {\bibfield
  {journal} {\bibinfo  {journal} {Phys. Rev. Lett.}\ }\textbf {\bibinfo
  {volume} {125}},\ \bibinfo {pages} {192002} (\bibinfo {year} {2020})},\
  \Eprint {http://arxiv.org/abs/2003.02288} {arXiv:2003.02288 [hep-ph]}
  \BibitemShut {NoStop}%
\bibitem [{\citenamefont {Gehrmann}\ \emph {et~al.}(2014)\citenamefont
  {Gehrmann}, \citenamefont {Luebbert},\ and\ \citenamefont
  {Yang}}]{Gehrmann:2014yya}%
  \BibitemOpen
  \bibfield  {author} {\bibinfo {author} {\bibfnamefont {T.}~\bibnamefont
  {Gehrmann}}, \bibinfo {author} {\bibfnamefont {T.}~\bibnamefont {Luebbert}},
  \ and\ \bibinfo {author} {\bibfnamefont {L.~L.}\ \bibnamefont {Yang}},\
  }\href {\doibase 10.1007/JHEP06(2014)155} {\bibfield  {journal} {\bibinfo
  {journal} {JHEP}\ }\textbf {\bibinfo {volume} {06}},\ \bibinfo {pages} {155}
  (\bibinfo {year} {2014})},\ \Eprint {http://arxiv.org/abs/1403.6451}
  {arXiv:1403.6451 [hep-ph]} \BibitemShut {NoStop}%
\bibitem [{\citenamefont {Echevarria}\ \emph
  {et~al.}(2016{\natexlab{a}})\citenamefont {Echevarria}, \citenamefont
  {Scimemi},\ and\ \citenamefont {Vladimirov}}]{Echevarria:2015usa}%
  \BibitemOpen
  \bibfield  {author} {\bibinfo {author} {\bibfnamefont {M.~G.}\ \bibnamefont
  {Echevarria}}, \bibinfo {author} {\bibfnamefont {I.}~\bibnamefont {Scimemi}},
  \ and\ \bibinfo {author} {\bibfnamefont {A.}~\bibnamefont {Vladimirov}},\
  }\href {\doibase 10.1103/PhysRevD.93.011502} {\bibfield  {journal} {\bibinfo
  {journal} {Phys. Rev. D}\ }\textbf {\bibinfo {volume} {93}},\ \bibinfo
  {pages} {011502} (\bibinfo {year} {2016}{\natexlab{a}})},\ \bibinfo {note}
  {[Erratum: Phys.Rev.D 94, 099904 (2016)]},\ \Eprint
  {http://arxiv.org/abs/1509.06392} {arXiv:1509.06392 [hep-ph]} \BibitemShut
  {NoStop}%
\bibitem [{\citenamefont {Echevarria}\ \emph
  {et~al.}(2016{\natexlab{b}})\citenamefont {Echevarria}, \citenamefont
  {Scimemi},\ and\ \citenamefont {Vladimirov}}]{Echevarria:2016scs}%
  \BibitemOpen
  \bibfield  {author} {\bibinfo {author} {\bibfnamefont {M.~G.}\ \bibnamefont
  {Echevarria}}, \bibinfo {author} {\bibfnamefont {I.}~\bibnamefont {Scimemi}},
  \ and\ \bibinfo {author} {\bibfnamefont {A.}~\bibnamefont {Vladimirov}},\
  }\href {\doibase 10.1007/JHEP09(2016)004} {\bibfield  {journal} {\bibinfo
  {journal} {JHEP}\ }\textbf {\bibinfo {volume} {09}},\ \bibinfo {pages} {004}
  (\bibinfo {year} {2016}{\natexlab{b}})},\ \Eprint
  {http://arxiv.org/abs/1604.07869} {arXiv:1604.07869 [hep-ph]} \BibitemShut
  {NoStop}%
\bibitem [{\citenamefont {Luo}\ \emph {et~al.}(2019)\citenamefont {Luo},
  \citenamefont {Wang}, \citenamefont {Xu}, \citenamefont {Yang}, \citenamefont
  {Yang},\ and\ \citenamefont {Zhu}}]{Luo:2019hmp}%
  \BibitemOpen
  \bibfield  {author} {\bibinfo {author} {\bibfnamefont {M.-X.}\ \bibnamefont
  {Luo}}, \bibinfo {author} {\bibfnamefont {X.}~\bibnamefont {Wang}}, \bibinfo
  {author} {\bibfnamefont {X.}~\bibnamefont {Xu}}, \bibinfo {author}
  {\bibfnamefont {L.~L.}\ \bibnamefont {Yang}}, \bibinfo {author}
  {\bibfnamefont {T.-Z.}\ \bibnamefont {Yang}}, \ and\ \bibinfo {author}
  {\bibfnamefont {H.~X.}\ \bibnamefont {Zhu}},\ }\href {\doibase
  10.1007/JHEP10(2019)083} {\bibfield  {journal} {\bibinfo  {journal} {JHEP}\
  }\textbf {\bibinfo {volume} {10}},\ \bibinfo {pages} {083} (\bibinfo {year}
  {2019})},\ \Eprint {http://arxiv.org/abs/1908.03831} {arXiv:1908.03831
  [hep-ph]} \BibitemShut {NoStop}%
\bibitem [{\citenamefont {Collins}\ and\ \citenamefont
  {Rogers}(2015)}]{Collins:2014jpa}%
  \BibitemOpen
  \bibfield  {author} {\bibinfo {author} {\bibfnamefont {J.}~\bibnamefont
  {Collins}}\ and\ \bibinfo {author} {\bibfnamefont {T.}~\bibnamefont
  {Rogers}},\ }\href {\doibase 10.1103/PhysRevD.91.074020} {\bibfield
  {journal} {\bibinfo  {journal} {Phys. Rev. D}\ }\textbf {\bibinfo {volume}
  {91}},\ \bibinfo {pages} {074020} (\bibinfo {year} {2015})},\ \Eprint
  {http://arxiv.org/abs/1412.3820} {arXiv:1412.3820 [hep-ph]} \BibitemShut
  {NoStop}%
\bibitem [{\citenamefont {Baranov}\ \emph {et~al.}(2021)\citenamefont {Baranov}
  \emph {et~al.}}]{CASCADE:2021bxe}%
  \BibitemOpen
  \bibfield  {author} {\bibinfo {author} {\bibfnamefont {S.}~\bibnamefont
  {Baranov}} \emph {et~al.} (\bibinfo {collaboration} {CASCADE}),\ }\href
  {\doibase 10.1140/epjc/s10052-021-09203-8} {\bibfield  {journal} {\bibinfo
  {journal} {Eur. Phys. J. C}\ }\textbf {\bibinfo {volume} {81}},\ \bibinfo
  {pages} {425} (\bibinfo {year} {2021})},\ \Eprint
  {http://arxiv.org/abs/2101.10221} {arXiv:2101.10221 [hep-ph]} \BibitemShut
  {NoStop}%
\bibitem [{\citenamefont {Abdulov}\ \emph {et~al.}(2021)\citenamefont {Abdulov}
  \emph {et~al.}}]{Abdulov:2021ivr}%
  \BibitemOpen
  \bibfield  {author} {\bibinfo {author} {\bibfnamefont {N.~A.}\ \bibnamefont
  {Abdulov}} \emph {et~al.},\ }\href {\doibase 10.1140/epjc/s10052-021-09508-8}
  {\bibfield  {journal} {\bibinfo  {journal} {Eur. Phys. J. C}\ }\textbf
  {\bibinfo {volume} {81}},\ \bibinfo {pages} {752} (\bibinfo {year} {2021})},\
  \Eprint {http://arxiv.org/abs/2103.09741} {arXiv:2103.09741 [hep-ph]}
  \BibitemShut {NoStop}%
\end{thebibliography}%

\end{document}